\documentclass[conference]{IEEEtran}

\RequirePackage[normalem]{ulem} 
\RequirePackage{color}\definecolor{RED}{rgb}{1,0,0}\definecolor{BLUE}{rgb}{0,0,1} 

\RequirePackage{color}

\usepackage{tikz}
\usepackage[sumlimits,intlimits]{amsmath}
\usepackage{amsmath,amsfonts,amssymb}%
\usepackage{bm}%
\usepackage{graphicx,graphics}%
\usepackage{cases}%
\usepackage[noadjust]{cite}%
\usepackage{color}%
\usepackage{cite,url}
\usepackage{verbatim}
\usepackage{enumerate}
\usepackage{algorithm}
\usepackage{algorithmic}
\usepackage{balance}
\usepackage{multirow}
\usepackage{stfloats}
\usepackage{subfigure}
\usepackage{xspace}
\usepackage{smartref}
 \usepackage{epstopdf}

\begin{document}

\title{\vspace*{-14pt}Enhanced List-Based Group-Wise Overloaded Receiver with Application to Satellite Reception
\vspace{-0.2cm}\\
}
\author{\vspace*{0pt}
\authorblockN{Zohair Abu-Shaban$^\dag$, Bhavani Shankar Mysore R$^\dag$, Hani Mehrpouyan$^\ddag$ and Bj\"{o}rn Ottersten$^\dag$.\vspace*{-10pt}}
\IEEEauthorblockA{\\$^\dag$ SnT -- University of Luxembourg, Luxembourg.\\
$^\ddag$ Department of ECE\&CS, California State University, Bakersfield, CA, USA.\\
emails: zohair.abushaban@uni.lu, hani.mehr@ieee.org, bhavani.shankar@uni.lu, bjorn.ottersten@uni.lu.
\vspace{-10pt}
}}
\maketitle
\thispagestyle{empty}

\begin{abstract}
The market trends towards the use of smaller dish antennas for TV satellite receivers, as well as the growing density of broadcasting satellites in orbit require the application of robust adjacent satellite interference (ASI) cancellation algorithms at the receivers. The wider beamwidth of a small size dish and the growing number of satellites in orbit impose an overloaded scenario, i.e., a scenario where the number of transmitting satellites exceeds the number of receiving antennas. For such a scenario, we present a two stage receiver to enhance signal detection from the satellite of interest, i.e., the satellite that the dish is pointing to, while reducing interference from neighboring satellites. Towards this objective, we propose an enhanced List-based Group-wise Search Detection (LGSD) receiver architecture that takes into account the spatially correlated additive noise and uses the signal-to-interference-plus-noise ratio (SINR) maximization criterion to improve detection performance. Simulations show that the proposed receiver structure enhances the performance of satellite systems in the presence of ASI when compared to existing methods.

\vspace{+3pt}

\end{abstract}

\section{Introduction}
{\let\thefootnote\relax\footnotetext{{
This work is supported by the National Research Fund (FNR), Luxembourg. Project ID: 4043055.
\vspace{-3pt}
}} }
Satellite Direct-to-Home (DTH) television broadcast services is a growing sector of satellite business and will continue to be a key driver for the overall satellite industry in the future \cite{SatIndustry}. To keep up with market demands, more satellites are launched and stationed in the geostationary orbit (GEO) causing the relevant frequency bands, e.g., Ku band, to be densely occupied. This increases the receiver susceptibility to \textit{adjacent satellite interference} (ASI) arising from neighboring satellites \cite{Elbert2004}. Furthermore, a smaller antenna size at the receiver is commercially attractive to home users as it reduces the manufacturing and mounting costs. However, smaller dishes are less directive and have wider reception beams which can result in a higher level of ASI at the receiver. The orbit occupancy and the small receive antenna size make ASI cancellation an urging priority to enable further future growth of the DTH business.

\begin{figure}[!t]
\begin{center}
  \includegraphics[scale=.75]{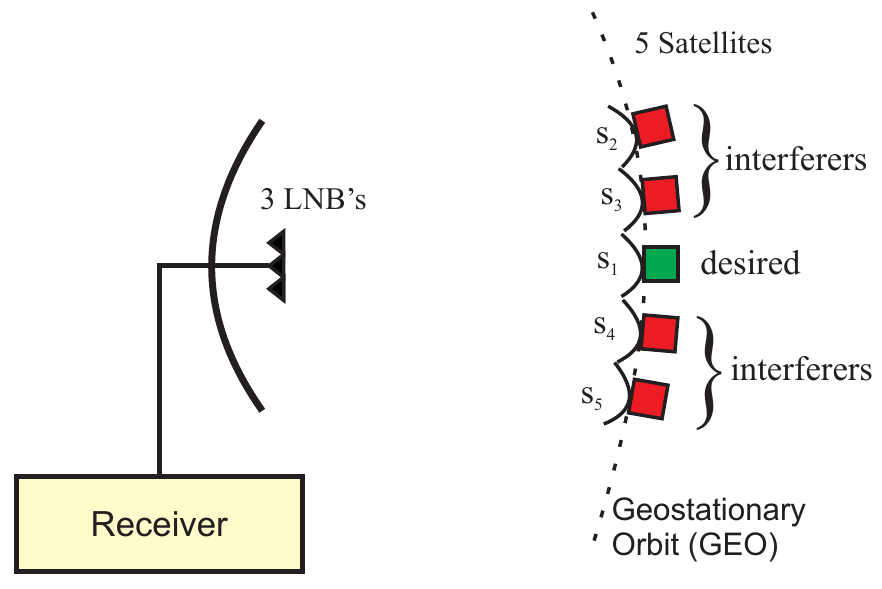}\\
  \caption{The system setup for 5 satellites and 3 LNBs. The dish is directed to the central or desired satellite, $s_1$.}\label{fig:setup}
  \end{center}
  \vspace{-.7cm}
\end{figure}

A multi-antenna satellite receiver dish employs multiple feeds, known as low noise blocks (LNBs). The number of LNBs should be kept low, e.g., 2-3 LNBs, due to cost, mechanical support and electromagnetic blockage issues \cite{grotz2010}. The increased number of satellites in view and the limited number of receiving LNBs motivate the consideration of overloaded receivers, receivers with fewer LNBs than received co-channel signals, e.g., see Fig. \ref{fig:setup}.

For overloaded receivers, multi-user detection and interference cancellation techniques are addressed in \cite{grotz2010,AbuShaban2012,krause2011,hicks2001,kapur2003,Colman2008,Bayram2000,Grant1998}. Interference cancellation techniques for both coded and uncoded signals with partial frequency overlapping are reported in \cite{beidas2002} and extended in \cite{Schwarz2007} to support digital video broadcasting via satellites standards, DVB-S and DVB-S2 \cite{dvbb-s2}. However, these two works do not exploit the spatial properties of the received signals.

In \cite{grotz2010}, by applying multiple LNBs (MLNBs), a two-stage multi-antenna receiver for satellite reception that is composed of a linear preprocessor stage and an iterative non-linear stage is presented. However, the work in \cite{grotz2010} is based on the assumption that the transmitted signals from the different satellites are only partially overlapping in frequency. Considering fully frequency overlapping transmitted signals, the approach in \cite{AbuShaban2012} employs successive interference cancellation (SIC) with a hybrid beamforming scheme to detect multiple satellites in an overloaded scenario. This allows different satellites to have different beamformers that best suit their spatial conditions. However, as shown in Section \ref{sec:sim}, the receiver in \cite{AbuShaban2012} fails to perform well when high order modulations are utilized by the satellite system.

As shown in \cite{Bayram2000}, the joint maximum likelihood (JML) detector is optimum for decoding co-channel signals in an overloaded system. However, its complexity grows exponentially with the number of transmitted signals. An attractive sub-optimum lower-complexity technique is the \textit{list-based group search detection} (LGSD) reported in \cite{krause2011}. An LGSD-based two-stage overloaded receiver employs a low-complexity search-based algorithm. The algorithm uses the maximum likelihood (ML) criterion to search within a lower-dimension signal space by splitting the received vector into a group of sub-vectors. The first stage of LGSD is a \textit{linear preprocessor} that contains a diversity combing scheme (beamformer) and a noise whitening filter. The maximum ratio combining (MRC) beamformer in \cite{Brennan1959} combines the received signals by maximizing the signal-to-noise ratio (SNR). Moreover, since white noise is required by the demodulator, a whitening filter is applied at the receiver. Compared to the optimum detection method, LGSD performs well in terms of bit error rate (BER) while also reducing complexity. However, in LGSD, interference is modeled as a white Gaussian process for diversity combining and the additive channel noise is assumed uncorrelated. These two assumptions may not hold for the satellite reception scenarios considered in the sequel. Thus, here, we focus on enhancing the performance of LGSD by addressing the first stage and modifying these two assumptions to suit the considered scenario.

In this paper, we tackle the design problem of an overloaded multi-antenna receiver with particular application to satellite broadcast reception. The receiver is equipped with a small aperture antenna, e.g., $<$40 cm, that has multiple LNBs. As shown in Fig. \ref{fig:setup}, the dish is assumed to be fixed and directed towards the central satellite, which we refer to as the \textit{desired} satellite. Other satellites operating in the same frequency band in view are referred to as \textit{interferers}. Due to the small dish size, the antenna patterns are wide, resulting in a high level of interference. In contrast to the scenario in \cite{krause2011}, the considered scenario assumes spatially correlated noise since the radiation patterns of the MLNB�s overlap causing one LNB noise pattern to affect the neighboring LNBs \cite{grotz2010}. We modify the LGSD receiver presented in \cite{krause2011} by proposing a preprocessor based on the signal-to-interference-and-noise (SINR) criterion and deriving a noise whitening filter. The performance of the receiver is measured in terms of BER and is compared with \cite{AbuShaban2012} and \cite{krause2011}. The contributions of this paper are summarized below:
\begin{itemize}
  \item Contrary to \cite{krause2011}, we use the spatial knowledge and the fixed antenna setup to accurately model the interference instead of treating it as additive noise at the receiver. Hence, a beamformer based on the SINR maximization criterion, i.e., the Wiener-Hopf beamformer, is utilized \cite{VanTrees}.
  \item Due to the antenna pattern overlap discussed above and atmospheric effects, a practical model of the additive noise that takes into account the correlation amongst the LNBs is considered here. Thus, a new whitening filter is derived that is better suited to the proposed beamformer and the more accurate model of the additive noise.
  \item By using the proposed beamforming scheme a new receiver structure denoted by \emph{Enhanced-LGSD} is proposed that can be applied in an overloaded satellite reception scenario to detect the desired satellite's signal. Extensive Monte-Carlo simulations are carried out to illustrate the performance of this receiver for both coded and uncoded scenarios.
\end{itemize}

Throughout this paper, we use the following notations: a scalar is denoted by an italic lowercase letter, while a column vector and a matrix are denoted by bold lowercase and uppercase letters, respectively. $\mathbf{I}_N$ denotes $N\times{N}$ identity matrix. $\|\mathbf{a}\|$ and $\|\mathbf{A}\|_F$ denote the  Euclidean norm of vector $\mathbf{a}$ and the Frobenius norm of matrix $\mathbf{A}$, respectively. For the transpose, the Hermitian, and pseudo-inverse operators, $(\cdot)^T$, $(\cdot)^H$, and $(\cdot)^\dag$ are used respectively. $\mathcal{C}^N$  refers to the $N$-dimension complex space. Greek letters are used to denote sets and subsets.   $\phi$ is the empty set while $\mid{\Gamma}\mid$ is the cardinality of set $\Gamma$. Finally, $\textit{a}^\circ$ is the angle $a$ in degrees.

The remainder of this paper is organized as follows: Section \ref{sec:section-sys-model} highlights the system model, the considered scenario, and the assumptions. Section \ref{sec:section-prop_rcv} briefly describing the LGSD detector and outlines the proposed preprocessor including the beamformer and the noise whitening filter. The simulation environment and results are discussed in Section \ref{sec:sim} while Section \ref{sec:conc} concludes the paper.

\section{System Model and Assumptions}\label{sec:section-sys-model}
We consider $N$ adjacent satellites stationed in GEO and broadcasting to an overloaded receiver connected to a small-size dish with $M$ LNBs. The following assumptions are made throughout this paper:
\begin{itemize}
  \item \emph{$N>M$}: Due to practical reasons such as cost reduction and electromagnetic blockage, the number of LNBs, $M$, is to be kept small i.e., 2-3 LNBs. For small-aperture reflectors, a larger number of satellites fall in the field of view of the antenna. Depending on the dish diameter, $D$, and the wavelength $\lambda$, the reflector $3$-dB beamwidth can be estimated by $(70\lambda/D)^\circ$ \cite{SatPrinciples}. The number of satellites can then be estimated knowing that the GEO satellites are usually separated by an angular spacing of $2.5^\circ-3^\circ$ \cite{grotz2010}. For example, the $3$-dB beamwidth of the central LNB of a dish with a diameter of $35$ cm operating in the Ku-band is $5^\circ-6^\circ$. Thus, one can expect $3$ satellites to fall within the field of view of the dish. Adding more LNBs extends the field of view and more satellites can be observed (See \cite{AbuShaban2012}).
  \item \emph{The transmitted signals from different satellites are assumed to occupy the same frequency band:} looking into the future, the orbital slots could be populate with co-channel satellites since both orbit and spectrum are scarce. ASI cancellation is achieved at the home receiver that uses a smaller size dish. This is a desirable situation where satellite broadcast service can grow and the receivers remain cost-effective.
  \item \emph{The system is synchronized as in \cite{AbuShaban2012} and \cite{Grotz2005}:} The LNBs can use the same oscillator to reduce the frequency uncertainty. However, the signals arrive at the receiver at different times. The synchronization parameters are assumed to be supplied by a synchronizer block at the digital front-end of the receiver. Moreover, when the satellites are operated by single operator, a better degree of synchronization can be expected.
  \item \emph{The additive noise is assumed to be spatially correlated:} The radiation patterns of the MLNBs overlap causing one LNB's noise pattern to affect the neighboring LNBs \cite{grotz2010}. This radiation overlap also correlates the noise emanating from other sources such as the gateway and satellite components.
  \item \emph{The signals comply with the DVB-S2 standard and are independently transmitted:} Signal parameters such as modulation, code rate, and power level are estimated by the receiver using the frame structure of DVB-S2.
  \item \emph{The channel is known and fixed:} A line-of-sight link and a clear sky are assumed. Therefore, the channel mainly depends on the antenna geometry and electrical specifications such as diameter, focal length, oscillator stability, low noise amplifier gain, etc. Since these parameters do not change quickly, they are assumed fixed over the transmission interval. Accordingly, ignoring pointing errors, the antenna radiation patterns are considered known and fixed.
\end{itemize}
Under the above assumptions, the received signal vector at the output of the synchronizer is modeled as
\begin{align}\label{eq:sys-eq}
\mathbf{r}[k] = \mathbf{As}[k]+\mathbf{n}[k],
\end{align}
where $\mathbf{r}[k]\triangleq\big[{r}_1 [k], {r}_2 [k], ..., {r}_M [k]\big]^T$ is the received symbol vector at time instant $k$, $\mathbf{A}\triangleq[A_{i,j}]$ is an $M\times{N}$ matrix representing the antenna array response with $A_{i,j}$ denoting the complex gain of the $i^{th}$ LNB in the direction of the $j^{th}$ satellite. Moreover, $\mathbf{s}[k]\triangleq\big[s_1 [k], s_2 [k], ..., s_{N}[k]\big]^T$ is the transmitted symbol vector where $s_j [k]$ is drawn from a zero-mean unit-variance signal constellation, $\omega$, $s_1$ corresponds to the desired satellite shown in Fig. \ref{fig:setup}, and $\mathbf{n}[k]\triangleq\big[n_1[k], n_2 [k], ..., n_M [k]\big]^T$ is the additive noise vector that is modeled as a Gaussian process with covariance matrix $\mathbf{R}_{nn}=\sigma_n^2\mathbf{K}$. Here, $\sigma_n^2$ is the noise power and $\mathbf{K}$ is the spatial correlation matrix.

\vspace{+5pt}
\section{The Proposed Receiver Design}\label{sec:section-prop_rcv}
\begin{figure}[!t]
\begin{center}
\vspace{-0.2cm}
  \includegraphics[scale=0.45]{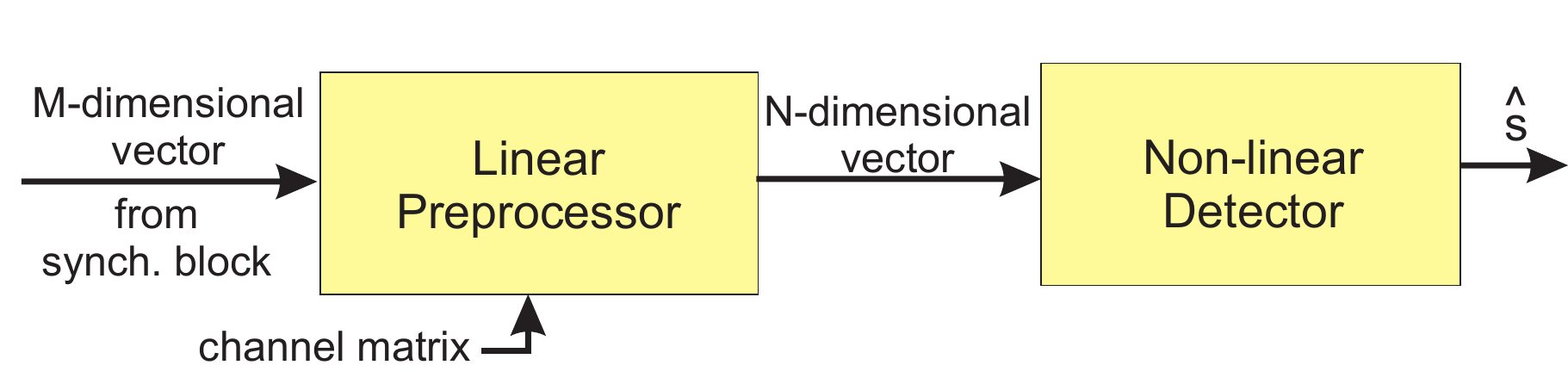}\\
    \vspace{-0.4cm}
  \caption{Overloaded receiver generic block diagram. }\label{fig:blcok_diagram}
  \end{center}
  \vspace{-0.7cm}
\end{figure}

 A generic block diagram for an overloaded receiver is shown in Fig. \ref{fig:blcok_diagram}. It is composed of two stages: the first stage is a \textit{linear preprocessor }comprising a beamformer and a noise whitening filter, while the second stage is a \textit{non-linear detector} that can be JML, LGSD, or the Enhanced-LGSD. The output is an estimated vector, $\hat{\mathbf{s}}$, of the transmitted symbols, $\mathbf{s}$. As shown in \cite{Bayram2000}, the JML detector for overloaded systems.

\subsection{Linear Preprocessors}\label{subsec:lpp}
Denoting by $\mathbf{W}$ and $\mathbf{F}$ the $M\times{N}$ beamforming matrix and the $N\times{N}$ noise whitening filter, respectively, and omitting the time index, $k$, the output of the linear preprocessor is given by
\begin{align}\label{eq:new-sys_eq}
\mathbf{y}=\mathbf{Hs}+\mathbf{z},
\end{align}
where $\mathbf{H}\triangleq\mathbf{F}^H\mathbf{W}^H\mathbf{A}$ is the equivalent channel matrix of size $N\times{N}$ and $\mathbf{z}\triangleq\mathbf{F}^H\mathbf{W}^H\mathbf{n}$ is the whitened Gaussian noise vector.

In the following we briefly mention the MRC beamformer used in \cite{krause2011} and then describe the proposed beamformer that improves upon this approach by maximizing the SINR instead.
\subsubsection{Maximum Ratio Combining (MRC)}\label{subsubsec:MRC_SNR}
Unlike the approach here, the preprocessor stage of the LGSD receiver in \cite{krause2011} is based on the assumption that the additive noise vector, $\mathbf{n}$ can be modeled as a white Gaussian process. Moreover, in \cite{krause2011}, the MRC combining scheme applied at the receiver is based on the SNR maximization criterion, i.e.,
\begin{align}
\mathbf{W} = \mathbf{A}.
\end{align}
Thus, it follows from \cite{krause2011} that
\begin{align}\label{eq:F-special}
\mathbf{F} = ((\mathbf{A}^H\mathbf{A})^\dag)^{\frac{1}{2}}.
\end{align}
As shown in Section \ref{subsec:uncoded}, based on the above design criteria, the LGSD receiver in \cite{krause2011} does not perform well in the overloaded satellite scenarios considered.

\subsubsection{Weiner-Hopf beamforming}\label{subsubsec:MRC_SINR}
Unlike the MRC approach in \cite{krause2011}, we propose a beamformer that accounts for the interference in the diversity combining and uses a preprocessor that maximizes the SINR.
Letting $\mathbf{A}\triangleq[\mathbf{a}_1,\mathbf{a}_2,...\mathbf{a}_{N}]$, we can rewrite (\ref{eq:sys-eq}) as
\begin{align}\label{eq:receive}
 \mathbf{r}=\sum_{m=1}^{N}\mathbf{a}_{m}s_{m}+\mathbf{n}.
\end{align} \vspace{-1pt}
Assuming that the transmitted signals, $s_m, \forall{m}$, are uncorrelated, the auto-covariance matrix for the received signal, $\mathbf{R}$, is given by
\begin{align}\label{eq:autocovariance_matrix}
 \mathbf{R}=\sum_{m=1}^{N}\mathbf{a}_{m}\mathbf{a}_{m}^H+\mathbf{R}_{nn}=\sum_{m=1}^{N}\mathbf{R}_m+\mathbf{R}_{nn}.
\end{align}
Accordingly, the beamformer that maximizes the SINR for the $m^{th}$ stream is given by \cite{AbuShaban2012}, \cite{grotz2010}
\begin{align}\label{eq:MRC_BF}
 \mathbf{w}_m \triangleq arg \max_{\mathbf{w}\in{\mathcal{C}^M}}\frac{\mathbf{w}^H\mathbf{R}_m\mathbf{w}}{\mathbf{w}^H(\mathbf{R}-\mathbf{R}_m)\mathbf{w}}, 1\leq{m}\leq{N}.
\end{align}
The solution of this generalized Rayleigh quotient is obtained by solving a generalized eigenvalue problem. Thus, $\mathbf{w}_m$ is the eigenvector corresponding to the largest eigenvalue of $(\mathbf{R}-\mathbf{R}_m)^{-1} \mathbf{R}_m$ \cite{eigen}. Subsequently, it can be shown that $\mathbf{w}_m$ corresponds to the well-known Wiener-Hopf beamformer given by \cite{VanTrees}
\begin{align}
\mathbf{w}_m=\mathbf{R}^{-1}\mathbf{a}_m,
\end{align}
and
\begin{align}
\mathbf{W}\triangleq[\mathbf{w}_1,\mathbf{w}_2,...\mathbf{w}_{N}]=\mathbf{R}^{-1}\mathbf{A}.
\end{align}

We now derive a whitening filter accounting for the spatially correlated noise. This is motivated by the fact that the demodulator requires $\mathbf{z}$ in (\ref{eq:new-sys_eq}) to be white. Since the covariance matrix of $\mathbf{z}$ that is given by
\begin{align}\label{eq:Rz_long}
\mathbf{R}_{zz}=\sigma_n^2\mathbf{F}^H\mathbf{W}^H\mathbf{K}\mathbf{W}\mathbf{F},
\end{align}
is rank deficient, we design the whitening filter $\mathbf{F}$ in order to minimize $\|\mathbf{F}^H\mathbf{G}\mathbf{F}-\mathbf{I}_N\|_F$, where $\mathbf{G}=\mathbf{W}^H\mathbf{K}\mathbf{W}$. Since $\mathbf{K}$ is a correlation matrix, $\mathbf{G}$ is positive semi-definite and its singular value decomposition is $\mathbf{G}=\mathbf{U}\mathbf{L}\mathbf{U}^H$. As a result, it is straight forward that a solution for $\mathbf{F}$ can be obtained as
\begin{align}\label{eq:F-genral}
\mathbf{F}=\mathbf{U}(\mathbf{L}^\dag)^\frac{1}{2},
\end{align}
where $\mathbf{L}$ is a diagonal matrix containing the eigenvalues of $\mathbf{G}$, and $\mathbf{U}$ is the matrix of eigenvectors of $\mathbf{G}$. Compared to \cite{krause2011}, the two filter structures are equivalent from a detection point of view if the complete received vector is taken into account.\footnote{$\|\mathbf{H}\|=\|\mathbf{U}\mathbf{H}\|$ for any unitary $\mathbf{U}$.} However, since LGSD uses subvectors in the decision process, the two approaches may not be equivalent. This will be made evident in the sequel.

\subsection{List Group Search Detection (LGSD)}\label{subsec:lpp}
A JML detector is an exhaustive search detector that selects a symbol vector from the \textit{N}-dimensional signal space, $\Omega=\omega^N$, by minimizing the Euclidean distance.
\begin{align}
\hat{\mathbf{s}}=arg \min_{\mathbf{s}\in{\Omega}} {\|\mathbf{y}-\mathbf{H}\mathbf{s}\|}^2.
\end{align}
Even though the JML is optimum, its computational complexity grows exponentially with $N$. This motivates the use of suboptimal techniques, such as LGSD, that have reduced complexity.

\begin{figure}[!t]
\begin{center}
  \includegraphics[scale=0.45]{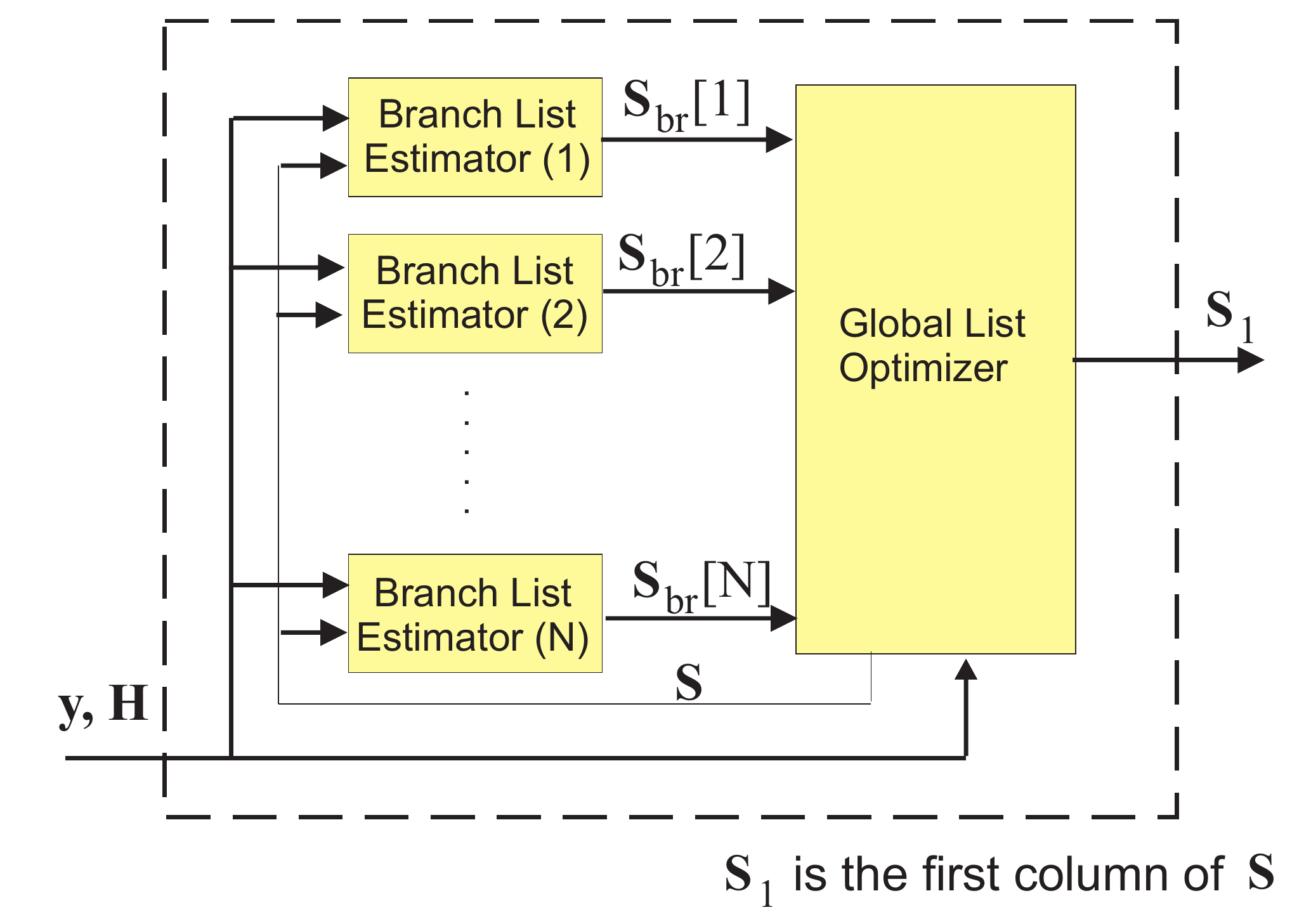}\\
      \vspace{-.4cm}
  \caption{Block diagram of the LGSD detector \cite{krause2011}.}\label{fig:LGSD}
  \end{center}
  \vspace{-0.7cm}
\end{figure}

The second stage of the proposed receiver is the LGSD detector depicted in Fig. \ref{fig:LGSD}.
The basic idea of LGSD is to split the transmit symbol vector into groups (subvectors) and perform an ML search over these shorter vectors. At the end, the results are combined to estimate the complete transmitted vector. Expressing $\mathbf{s}\triangleq\left[s_1, s_2, ..., s_{N}\right]^T \in\Omega=\omega^N$, let the index set $\Gamma=\{\gamma_1, \gamma_2,...\gamma_g,...\gamma_G\}$ such that $\gamma_i\cap\gamma_j=\phi$, $\forall i\neq{j}$, and $\cup_{g=1}^{G}\gamma_g=\{1,2,...,N\}=\Gamma$. We use the group index sets, $\gamma_g$, to map different $s_i$, for $1\leq{i}\leq{N}$, to $\mathbf{s}_g$ of size $\mid{\gamma_g}\mid\times{1}$ and to map the columns of $\mathbf{H}$ to $\mathbf{H}_g$ of size $N\times{\mid{\gamma_g}\mid}$. Subsequently, we can write (\ref{eq:new-sys_eq}) in terms of the groups $G$ as
\begin{align}\label{eq:group-sys_eq}
\mathbf{y}=\sum_{g=1}^{G}\mathbf{H}_g\mathbf{s}_g+\mathbf{z}.
\end{align}
To only detect group $j$ only, for $1\leq{j}\leq{G}$, we can write
\begin{align}
\mathbf{y}=\mathbf{H}_j\mathbf{s}_j+\sum_{g=1, g\neq{j}}^{G}\mathbf{H}_g\mathbf{s}_g+\mathbf{z},
\end{align}
\vspace{-0.2cm}
\begin{align}
\mathbf{y}_j\triangleq\mathbf{H}_j\mathbf{s}_j+\mathbf{z}=\mathbf{y}-\sum_{g=1, g\neq{j}}^{G}\mathbf{H}_g\mathbf{s}_g,
\end{align}
\vspace{-0.2cm}
\begin{align}\label{eq:group_ML}
\hat{\mathbf{s}}_j\triangleq arg \min_{\mathbf{s}_j\in{\omega^{\mid\gamma_j\mid}}} {\|\mathbf{y}_j-\mathbf{H}_j\mathbf{s}_j\|}^2,
\end{align}
where $\mathbf{H}_j$ and $\mathbf{s}_j$ are the columns of $\mathbf{H}$ and the rows of $\mathbf{s}$, respectively, whose indices are included in the set $\gamma_j$. This is a less complex search than JML, since $\mathbf{s}_j$ is shorter than $\mathbf{s}$. The LGSD detector has two sub-processes as shown in Fig. \ref{fig:LGSD}, the branch list estimator (BLE) process and the global list optimizer (GLO) process. The $n^{th}$ BLE operates over the $n^{th}$ row of the input $\mathbf{y}$ and $\mathbf{H}$ and applies (\ref{eq:group-sys_eq})$-$(\ref{eq:group_ML}). The output of the $n^{th}$ BLE is an $N\times{L}$ matrix $\mathbf{S}_{br}[n]=\{\mathbf{s}_{br}^{(l)}[n]\}$, for $1\leq{l}\leq{L}$, i.e., a \textit{list} of $L$ vectors that is sorted in ascending order using the mean-square error (MSE) criterion given by
\begin{align}\label{eq:mse_row}
e_l(n)=\big\|\mathbf{y}(n)-\mathbf{H}(n) \mathbf{s}_{br}^{(l)}[n]\big\|^2.
\end{align}
In \eqref{eq:mse_row}, $\mathbf{y}(n)$ and $\mathbf{H}(n)$ are the $n^{th}$ rows of $\mathbf{H}$ and $\mathbf{y}$, respectively, while $\mathbf{s}_{br}^{(l)}[n]$ is drawn from the output list $\mathbf{S}_{br}[n]$. Subsequently, the lists from the BLEs are optimized by the GLO. The GLO operates over the columns of the channel matrix and produces a new list of vectors that are, again, sorted by their MSE and the result is sent to a hard detector that selects the first vector. To further enhance the detection, $\Theta$ and $\Phi$ iterations are run by the BLE and the GLO, respectively. Moreover, the entire LGSD algorithm is executed for $Q$ iterations. Further details on the LGSD detector can be found in \cite{krause2011}.
\vspace{0pt}
\section{Simulation Results and Discussion}\label{sec:sim}
In this section, we investigate the performance of the proposed enhanced preprocessor in terms of BER.
\subsection{Setup}
Monte-Carlo simulations are carried out for both coded and uncoded 8 phase-shift keying (8PSK) signals and 16 amplitude and phase-shift keying (16APSK) signals \cite{dvbb-s2}. In satellite broadcasting, forward error correction codes are used. In the sequel, we apply low density parity check (LDPC) code with a rate of ($3/4$). The considered setup is depicted in Fig. \ref{fig:setup} and consists of $M=3$ LNBs at the receiver and $N=5$ GEO satellites stationed at $0^\circ$, $-5.9^\circ$, $-2.8^\circ$, $3^\circ$ and $5.7^\circ$. These angles are measured clockwise relative to the central satellite. For typical ASI scenario realization \cite{grotz2010}, the satellites are separated by an angular spacing of $2.7^\circ-3^\circ$ and the LNBs are assumed to be mounted on a 35-cm dish which is directed towards the central satellite, $s_1$, in Fig. \ref{fig:setup}. Given the dish size, the reflector antenna analysis software GRASP \cite{GRASP} is used to obtain the channel matrix. This software is widely used by satellite research and professional teams since it accurately models the characteristics of parabolic antennas and creates realistic antenna patterns. The noise at the receiver is assumed to be spatially correlated and its correlation matrix is given by \cite{AbuShaban2012}
\begin{align} \label{eq:K}
\mathbf{K}=
\left(
  \begin{array}{ccc}
    1& 0.1& 0.05\\0.1& 1 &0.1\\ 0.05& 0.1& 1\\
  \end{array}
\right).
\end{align}

The SNR is obtained in terms of the average received power per signal such that, SNR$=\frac{\|\textbf{A}\|_F^2}{\sigma_n^2MN}$. We focus on the central satellite, since the dish is directed toward it, and it has the highest interference. For this reason, only the BER results for the central satellite are shown here. The results for the LGSD and the Enhanced-LGSD are obtained using the following LGSD parameters \cite{krause2011}, $L$, $Q$, $\Theta$ and $\Phi$, corresponding to output list length, overall LGSD iterations, BLE iterations, and GLO iterations, respectively. We use $L=4N$ while the values of iterations parameters are noted on the figures using the notations $(Q/\Theta/\Phi)$. These parameters govern a performance-complexity/trade-off, as discussed in Section \ref{subsec:complexity}.

The received vector is divided into two index groups, $\gamma_1$ and $\gamma_2$, of sizes $\mid\gamma_1\mid=3$ and $\mid\gamma_2\mid=2$. This division is chosen to allocate the three strongest satellites, i.e., $ s_1, s_3$ and $s_4$ in Fig. \ref{fig:setup}, to $\gamma_1$ and the remaining satellites to $\gamma_2$ in the first iteration. This follows the methodology in \cite{krause2011}, where in the first iteration, the groups are allocated such that the 3 signals with highest powers are allocated to $\gamma_1$. In the subsequent iterations, the allocation is randomized. This division represents an acceptable trade-off between complexity and performance, since larger groups require searching over larger spaces, while by selecting smaller groups the advantages of joint processing diminishes.
\subsection{Results for Uncoded Signals}\label{subsec:uncoded}
\begin{figure}[!t]
\begin{center}
  \includegraphics[scale=0.61]{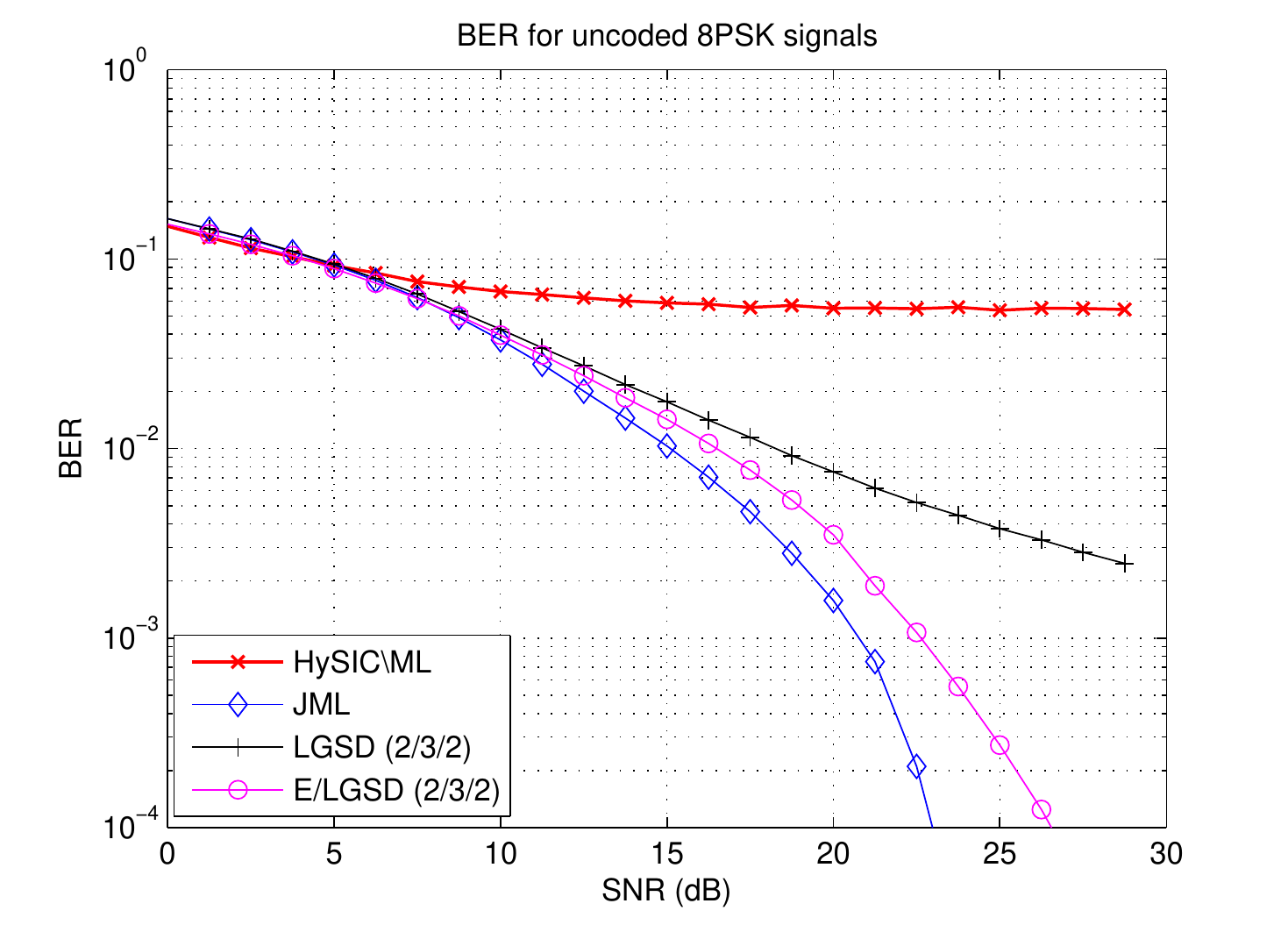}\\
  \vspace{-.3cm}
  \caption{Central satellite uncoded 8PSK BER of LGSD \cite{krause2011}, Enhanced-LGSD, and JML. Both LGSD and Enhanced-LGSD are obtained using iterations $(Q/\Theta/\Phi)=(2/3/2)$
 }\label{fig:uncoded8}
  \end{center}
  \vspace{-0.5cm}
\end{figure}
The BER curves for the different detection algorithms for 8PSK signals are presented in Fig. \ref{fig:uncoded8}. JML represents the lower bound BER performance. HySIC/ML \cite{AbuShaban2012} is a low-complexity approach that attempts to detect $s_1$ disjointly, after some preprocessing. Evidently, it does not perform very well in the case of 8PSK signals. Hence, joint processing algorithms, e.g., LGSD, that not only detect $s_1$ but also detect the interferers to enhance the overall system performance are preferred. By changing the linear preprocessor of LGSD, while maintaining the same number of iterations, Enhanced-LGSD reduces the gap with JML and improves the performance by some $7$ dB. This gain is achieved without any added complexity to LGSD. Moreover, Enhanced-LGSD moves the BER floor that is observed in the LGSD curve to a significantly lower value. Note that the curves corresponding to LGSD and Enhanced-LGSD in Fig. \ref{fig:uncoded8} are obtained using overall iteration $Q=2$, BLE iterations $\Theta=3$, and GLO iterations $\Phi=2$, i.e., (2/3/2).
\begin{figure}[!t]
\begin{center}
   \includegraphics[scale=0.63]{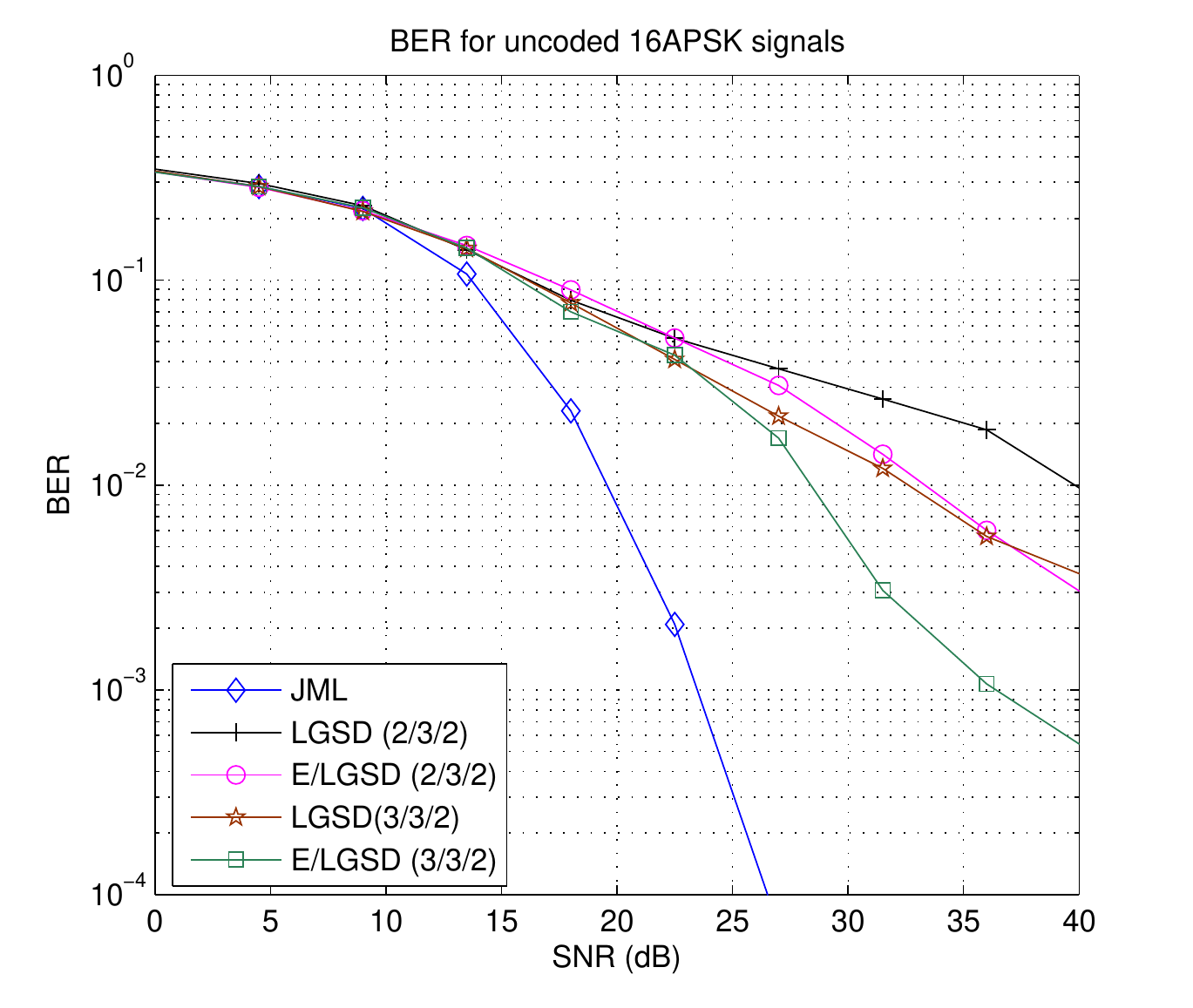}\\
  \vspace{-.3cm}
  \caption{Central satellite uncoded 16APSK BER of LGSD, Enhanced-LGSD and JML.
 }\label{fig:uncoded16}
  \end{center}
  \vspace{-.3cm}
\end{figure}
\begin{figure}[!t]
\begin{center}
 \includegraphics[scale=0.61]{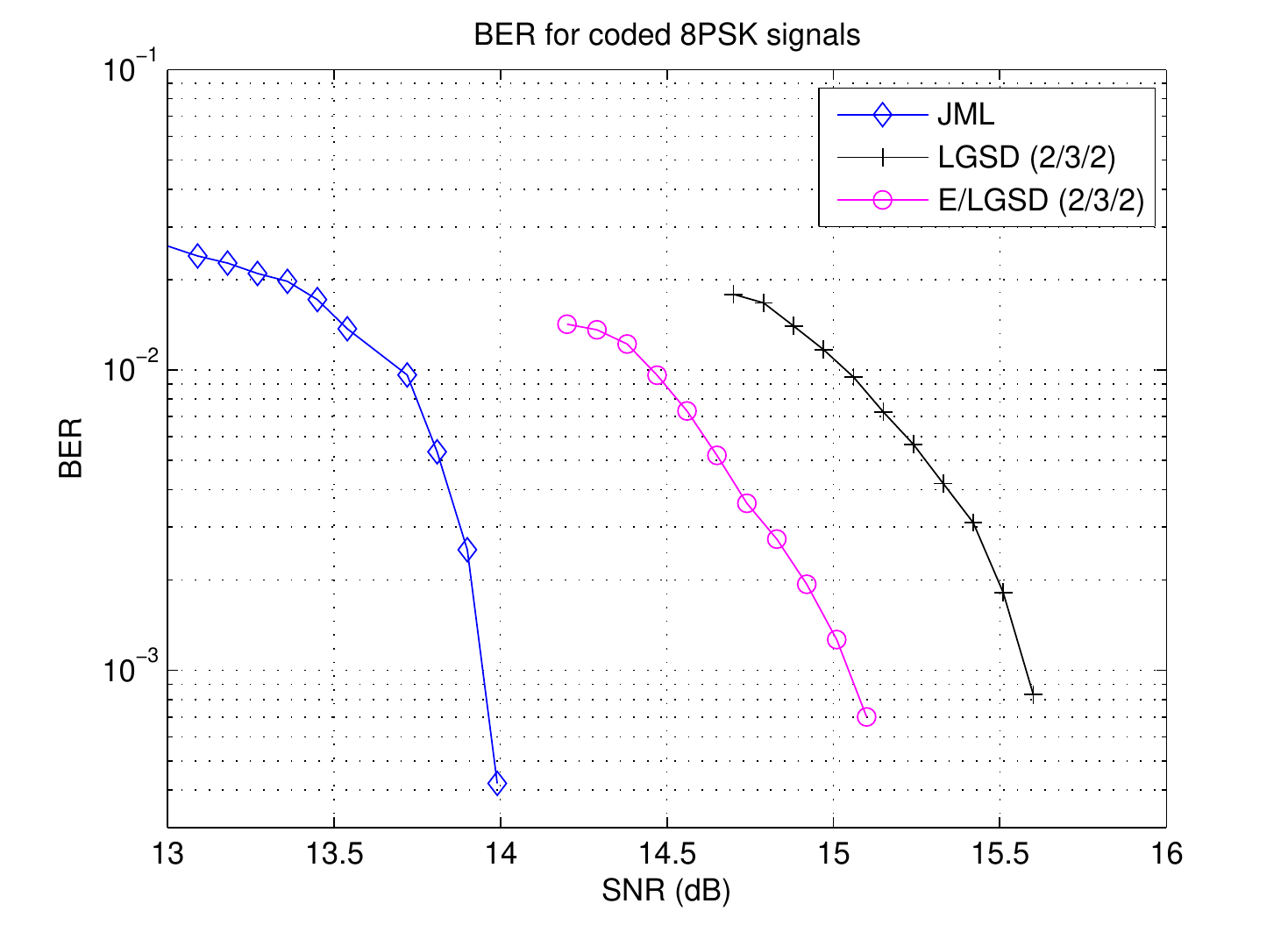}\\
  \vspace{-.3cm}
  \caption{Central satellite coded 8PSK BER of LGSD, Enhanced-LGSD and JML. Obtained using LDPC with a code rate of $3/4$.
}\label{fig:coded8}
  \end{center}
\vspace{-.7cm}
\end{figure}
The BER of HySIC/ML is not simulated for 16APSK due to its poor performance. Considering Fig. \ref{fig:uncoded16}, the performance of LGSD and Enhanced-LGSD are shown to diverge from JML performance. However, it can be observed that Enhanced-LGSD outperforms LGSD, when the same numbers of iterations are used. On the other hand, applying Enhanced-LGSD(3/3/2) instead of Enhanced-LGSD(2/3/2) results in a power gain of 9 dB at $3\times{10^{-3}}$ BER. Notice that LGSD(3/3/2) provides similar results as Enhanced-LGSD(2/3/2), although it uses higher number of iterations. With additional complexity, more iterations can be used to further enhance performance (see Section \ref{subsec:complexity}).
  \vspace{-0.1cm}
\subsection{Results for Coded Signals}\label{subsec:coded}
\begin{figure}[!t]
\begin{center}
 \includegraphics[scale=0.61]{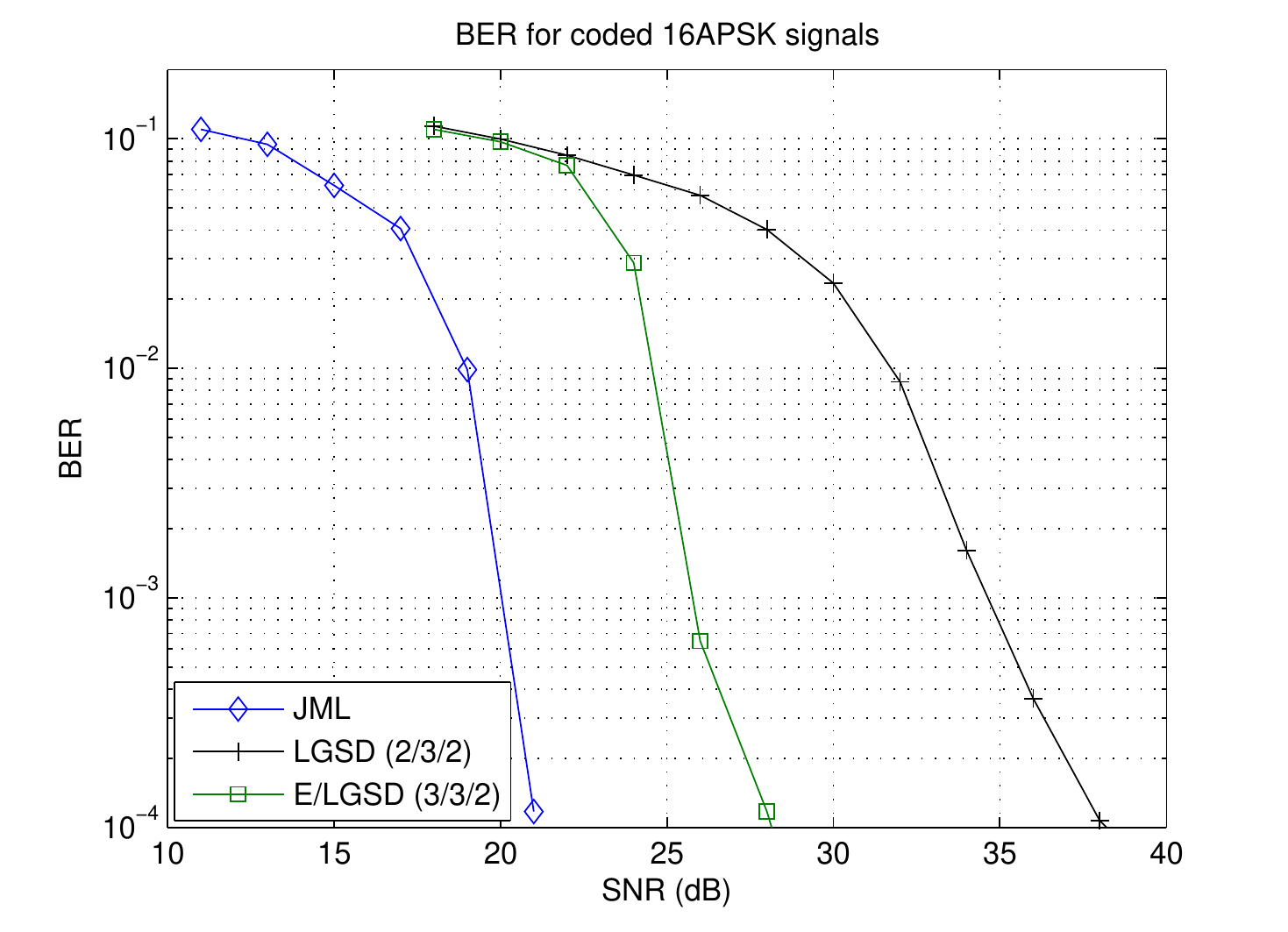}
  \vspace{-.5cm}
\caption{Central satellite coded 16APSK BER of LGSD, Enhanced-LGSD and JML. Obtained using LDPC with a code rate of $3/4$.
}
\label{fig:coded16}
  \end{center}
  \vspace{-0.8cm}
\end{figure}
The BER performances of different detectors for 8PSK coded signals are shown in Fig. \ref{fig:coded8}. Due to its poor performance, HySIC/ML is omitted from the coded simulations. There is a $0.5$ dB SNR gain by using the proposed preprocessor. Using a code rate of $3/4$, the gap between Enhanced-LGSD and JML is reduced to $1$ dB.

Fig. \ref{fig:coded16} illustrates the BER curves for coded transmission of 16APSK signals. Compared to the conventional LGSD, it is evident that using the proposed preprocessor, a gain of about $8$ dB is achieved, when using one extra overall iteration, i.e., Enhanced-LGSD(3/3/2). The gap with JML reduces to $6$ dB at a BER of $10^{-3}$.
  \vspace{-0.1cm}
\subsection{Complexity/Performance Analysis}\label{subsec:complexity}
We now discuss the effect of changing the number of iterations on both the performance and complexity of the proposed receiver. As in \cite{krause2011}, the complexity is measured by the number of squaring operations required to calculate the Euclidian distance metric. For the JML detector, the complexity is given by
\begin{align}
C_2=2N\mid\omega\mid^N.
\end{align}
The factor $2N$ is used since $N$ complex squaring operations are required for an $N\times{1}$ complex vector and two real squaring operations are required per entry. For both LGSD and Enhanced-LGSD, the complexity is obtained using the same approach as in \cite{krause2011}.
\begin{figure}[!t]
\begin{center}
 \includegraphics[scale=0.61]{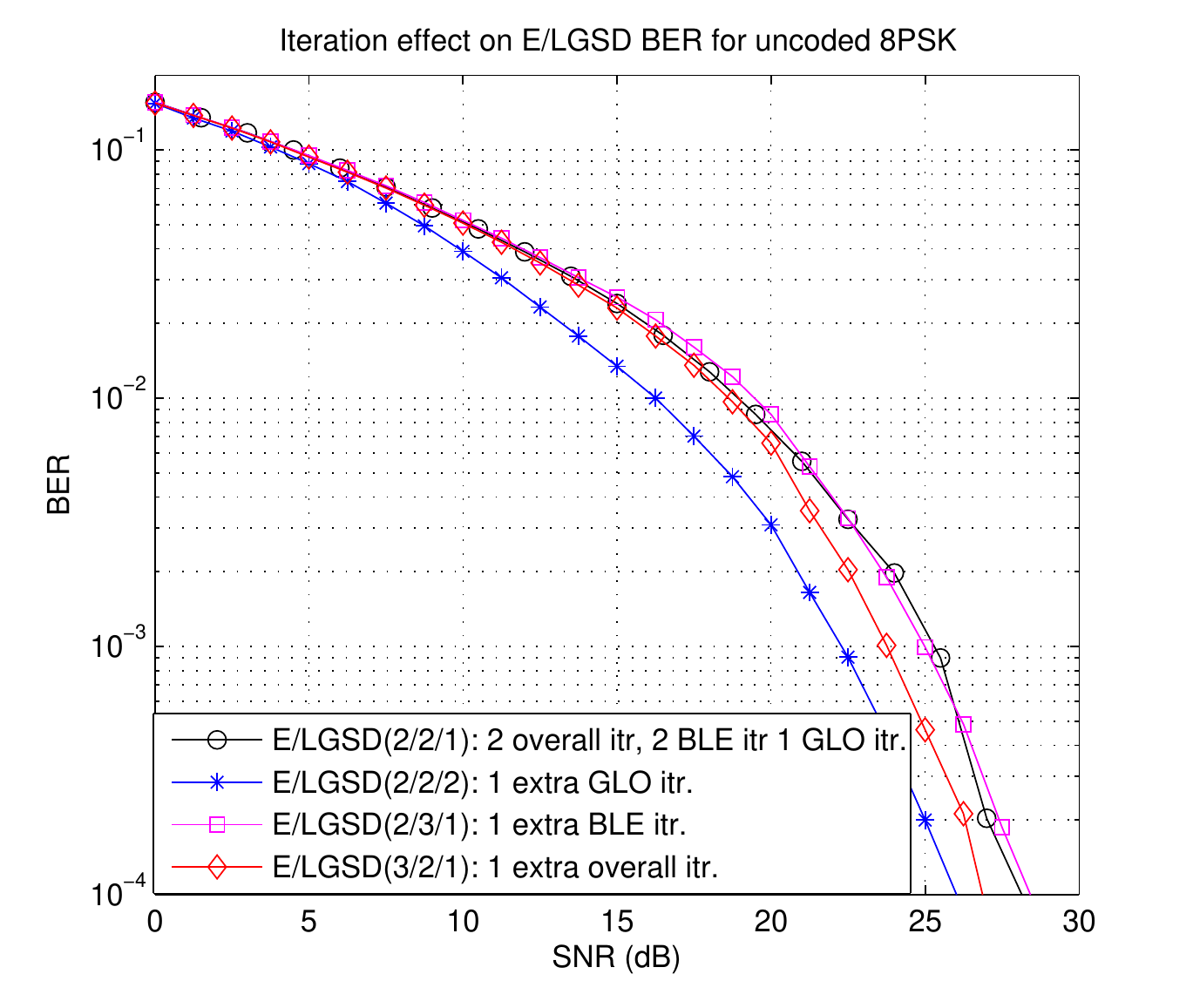}\\
  \vspace{-.3cm}
  \caption{Added complexity effect on BER performance for uncoded 8PSK signals. }\label{fig:complexity}
  \end{center}
\end{figure}
\begin{figure}[!t]
\begin{center}
  \vspace{-.3cm}
 \includegraphics[scale=0.61]{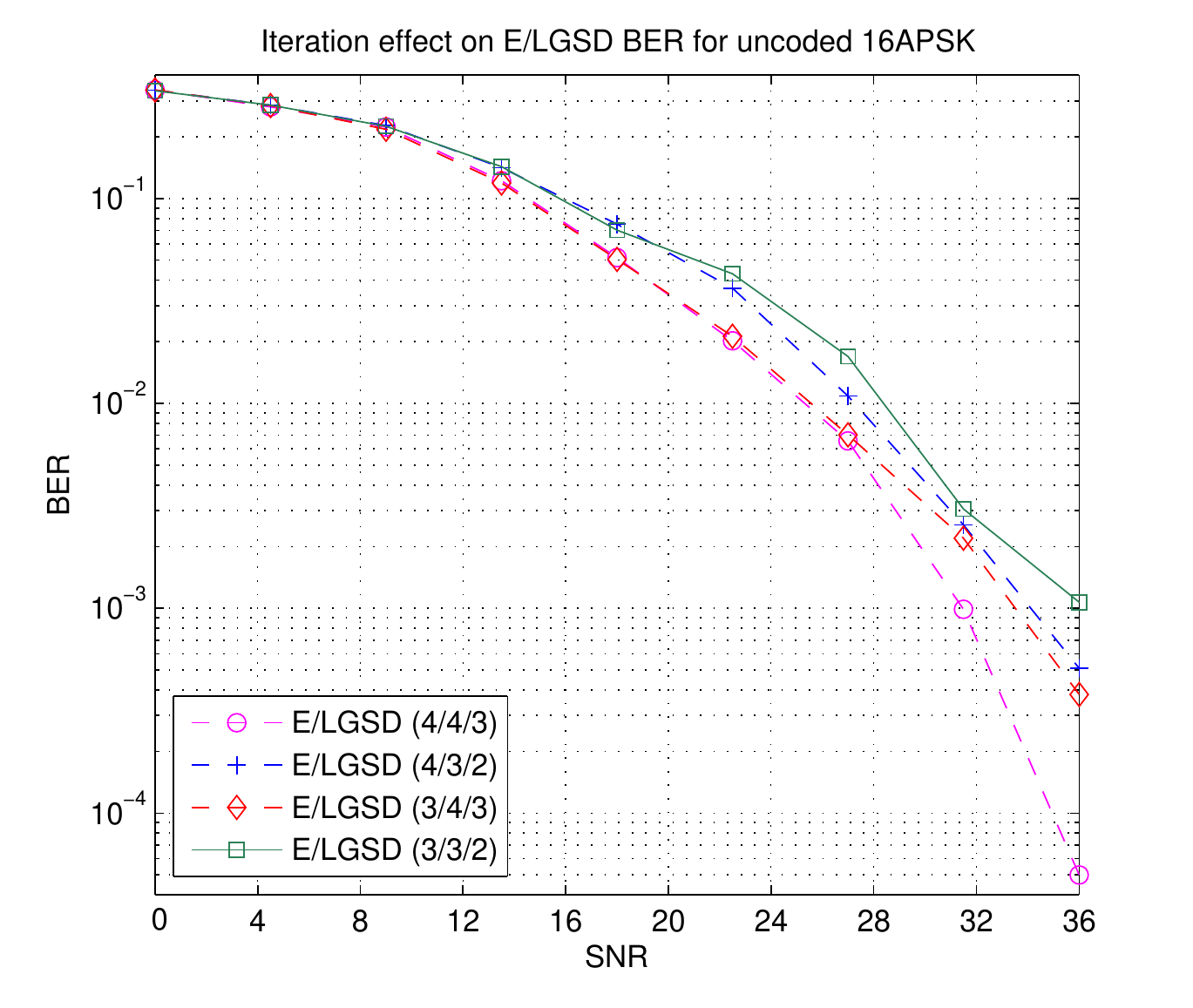}\\
  \vspace{-.3cm}
  \caption{Added complexity effect on BER performance for uncoded 16APSK signals. }\label{fig:complexity16}
  \end{center}
\vspace{-0.8cm}
\end{figure}
Measured as a percentage of the JML complexity, Table I summarizes the complexity of the simulation scenarios discussed in Figs. \ref{fig:uncoded8}, \ref{fig:uncoded16}, \ref{fig:complexity} and \ref{fig:complexity16}. Focusing on the effect of the number of iterations on the Enhanced-LGSD performance applied to 8PSK signals, it is observed that the number of iterations allocated to different stages of the detector should be selected more carefully. Compared to Enhanced-LGSD (2/2/1), adding one GLO iteration is equivalent to 28$\%$ of complexity increase and 3-4 dB of performance gain from Fig. \ref{fig:complexity}. However, adding one BLE iteration results in a complexity increase of 4$\%$ while providing almost no performance gain. In addition, in the case of 8PSK, the addition of an overall LGSD iteration also does not improve performance by a large margin, e.g., compared to Enhanced-LGSD (2/2/1), the complexity of Enhanced-LGSD (3/2/1) is $18\%$ higher while only providing $1$ dB performance gain. Indeed, one GLO iteration is expected to enhance the joint detection performance the since the whole symbol vector is involved in the detection process within the GLO iteration, while, in contrast, in a BLE iteration only individual subvectors of symbol are used in the detection process.

When considering $16$APSK signals in Fig. \ref{fig:uncoded16}, by applying Enhanced-LGSD (2/3/2) instead of Enhanced-LGSD (3/3/2), a performance gain of $9$ dB is obtained while the complexity is increased by $8\%$. With reference to Fig. \ref{fig:complexity16}, moving from Enhanced-LGSD(3/3/2) to Enhanced-LGSD(4/4/3) results in 5 dB of power gain while imposing 22\% of complexity increase on the detector. Even with this complexity increase, the complexity of E-LGSD(4/4/3) is 46\% of the JML complexity.

\vspace{-0pt}
\section{Conclusions}\label{sec:conc}
\begin{table}\label{tabel:complexity}
 \begin{center}
  \caption{Complexity of different detectors for different modulation orders}
\renewcommand{\arraystretch}{1.2}
\begin{tabular}{|l| c |l |c |}

  \hline
\multicolumn{2}{|c|}{8PSK} & \multicolumn{2}{|c|}{16APSK} \\
    \hline
JML & $3.30\times{10^5}$& JML& $10.4\times{10^6}$\\
\hline
LGSD(2/3/2)&  67\%&LGSD(2/3/2)& 16\%\\
\hline
Enh. LGSD(2/3/2)&  67\%& Enh. LGSD(2/3/2)&16\%\\
\hline
Enh. LGSD(2/2/1)&  35\%& Enh. LGSD(3/3/2)&24\%\\
\hline
Enh. LGSD(2/2/2)&  63\%& Enh. LGSD(3/4/3)&35\%\\
\hline
Enh. LGSD(2/3/1)&  39\%& Enh. LGSD(4/3/2)&32\%\\
\hline
Enh. LGSD(3/2/1)&  53\%& Enh. LGSD(4/4/3)&46\%\\
\hline
\end{tabular}
\end{center}
\vspace{-0.8cm}
\end{table}
This paper presents an enhanced-LGSD receiver that modifies the linear preprocessor in the conventional LGSD receiver. An SINR-based beamformer, known as the Wiener-Hopf beamformer, is used instead of the MRC approach. A whitening filter to account for the spatially correlated noise and the beamforming process is derived. The enhanced receiver is applied to satellite broadcast reception in an overloaded setup. Simulation results show that the receiver with the proposed linear preprocessor improves the performance in terms of BER. Although, no complexity reduction to LGSD is claimed, there are significant power savings depending on the considered modulation and complexity of the detector. It is well established in the satellite research community, that any power saving is crucial given the limited on-board power budget.

\vspace{-0pt}


\begin{thebibliography}{10}
\providecommand{\url}[1]{#1}
\csname url@samestyle\endcsname
\providecommand{\newblock}{\relax}
\providecommand{\bibinfo}[2]{#2}
\providecommand{\BIBentrySTDinterwordspacing}{\spaceskip=0pt\relax}
\providecommand{\BIBentryALTinterwordstretchfactor}{4}
\providecommand{\BIBentryALTinterwordspacing}{\spaceskip=\fontdimen2\font plus
\BIBentryALTinterwordstretchfactor\fontdimen3\font minus
  \fontdimen4\font\relax}
\providecommand{\BIBforeignlanguage}[2]{{%
\expandafter\ifx\csname l@#1\endcsname\relax
\typeout{** WARNING: IEEEtran.bst: No hyphenation pattern has been}%
\typeout{** loaded for the language `#1'. Using the pattern for}%
\typeout{** the default language instead.}%
\else
\language=\csname l@#1\endcsname
\fi
#2}}
\providecommand{\BIBdecl}{\relax}
\BIBdecl

\bibitem{SatIndustry}
\BIBentryALTinterwordspacing
{Satellite Industry Association}, ``{S}tate of the {S}atellite {I}ndustry
  {R}eport,'' June 2013. [Online]. Available: \url{http://www.sia.org/}
\BIBentrySTDinterwordspacing

\bibitem{Elbert2004}
B.~Elbert, \emph{Satellite Communication Applications Handbook}, 2nd~ed.\hskip
  1em plus 0.5em minus 0.4em\relax Artech House, Norwood, MA, USA, 2004.

\bibitem{grotz2010}
J.~Grotz, B.~Ottersten, and J.~Krause, ``Signal detection and synchronization
  for interference overloaded satellite broadcast reception,'' \emph{{IEEE}
  Trans. Wireless Commun.}, vol.~9, no.~10, pp. 3052 --3063, Oct. 2010.

\bibitem{AbuShaban2012}
Z.~Abu-Shaban, H.~Mehrpouyan, J.~Grotz, and B.~Ottersten, ``Overloaded
  satellite receiver using {SIC} with hybrid beamforming and {ML} detection,''
  in \emph{Proc. of 14th Workshop on Signal Processing Advances in Wirelss
  Commun. (SPAWC), Darmstadt, Germany}, June 2012, pp. 421 --425.

\bibitem{krause2011}
M.~Krause, D.~Taylor, and P.~Martin, ``List-based group-wise symbol detection
  for multiple signal communications,'' \emph{{IEEE} Trans. Wireless Commun.},
  vol.~10, no.~5, pp. 1636 --1644, May 2011.

\bibitem{hicks2001}
J.~Hicks, S.~Bayram, W.~Tranter, R.~Boyle, and J.~Reed, ``Overloaded array
  processing with spatially reduced search joint detection,'' \emph{{IEEE} J.
  Sel. Areas Commun.}, vol.~19, no.~8, pp. 1584 --1593, Aug 2001.

\bibitem{kapur2003}
A.~Kapur and M.~Varanasi, ``Multiuser detection for overloaded {CDMA}
  systems,'' \emph{{IEEE} Trans. Inf. Theory}, vol.~49, no.~7, pp. 1728 --
  1742, Jul. 2003.

\bibitem{Colman2008}
G.~Colman and T.~Willink, ``Overloaded array processing using genetic
  algorithms with soft-biased initialization,'' \emph{{IEEE} Trans. Veh.
  Technol.}, vol.~57, no.~4, pp. 2123 --2131, July 2008.

\bibitem{Bayram2000}
S.~Bayram, J.~Hicks, R.~Boyle, and J.~Reed, ``Joint maximum likelihood approach
  in overloaded array processing,'' in \emph{Proc. Vehicular Tech. Conf. Tokyo,
  Japan}, vol.~1, 2000, pp. 394--400.

\bibitem{Grant1998}
S.~Grant and J.~Cavers, ``Performance enhancement through joint detection of
  cochannel signals using diversity arrays,'' \emph{IEEE Transactions on
  Communications}, vol.~46, no.~8, pp. 1038--1049, Aug. 1998.

\bibitem{beidas2002}
B.~Beidas, H.~El~Gamal, and S.~Kay, ``Iterative interference cancellation for
  high spectral efficiency satellite communications,'' \emph{{IEEE} Trans.
  Commun.}, vol.~50, no.~1, pp. 31 --36, Jan. 2002.

\bibitem{Schwarz2007}
K.~Schwarzenbarth, J.~Grotz, and B.~Ottersten, ``{MMSE} based interference
  processing for satellite broadcast reception,'' in \emph{Proc. Vehicular
  Tech. Conf., Dublin, Ireland}, Apr. 2007, pp. 1345 --1349.

\bibitem{dvbb-s2}
{The European Telecommunications Standards Institute}, ``{EN} 302 307:
  {D}igital {V}ideo {B}roadcasting {(DVB)}; second generation framing
  structure, channel coding and modulation systems for broadcast, interactive
  services, news gathering and other broadband satellite applications,''
  \emph{ETSI}, Feb. 2005.

\bibitem{Brennan1959}
D.~G. Brennan, ``Linear diversity combining techniques,'' \emph{Proceedings of
  the Institute of Radio Engineers}, 1959.

\bibitem{VanTrees}
H.~L. Van~Trees, \emph{Adaptive Beamformers}.\hskip 1em plus 0.5em minus
  0.4em\relax John Wiley \& Sons, Inc. New York, USA, 2002, pp. 710--916.

\bibitem{SatPrinciples}
A.~K. Maini and V.~Agrawal, \emph{Satellite Technology: Principles and
  Applications}, 2nd~ed.\hskip 1em plus 0.5em minus 0.4em\relax John Wiley \&
  Sons, Chichester, UK, Oct. 2010.

\bibitem{Grotz2005}
J.~Grotz, J.~Krause, and B.~Ottersten, ``Decision-directed interference
  cancellation applied to satellite broadcast reception,'' in \emph{Proc.
  Vehicular Tech. Conf. Dallas, USA}, 2005.

\bibitem{eigen}
T.~De~Bie, N.~Cristianini, and R.~Rosipal, ``Eigenproblems in pattern
  recognition,'' in \emph{Handbook of Geometric Computing: Applications in
  Pattern Recognition, Computer Vision, Neural computing, and Robotics}.\hskip
  1em plus 0.5em minus 0.4em\relax E. B. Corrochano, ed. Berlin Heidelberg,
  Germany: Springer, 2005, pp. 128--132.

\bibitem{GRASP}
\BIBentryALTinterwordspacing
Ticra, ``{GRASP$^\circledR$} student edition 10.0.1,'' 2013. [Online].
  Available: \url{http://www.ticra.com/}
\BIBentrySTDinterwordspacing

\end{thebibliography}
\end{document}